\documentclass[%
 reprint,
 amsmath,amssymb,
 aps,
prb,
]{revtex4-2}
\usepackage{graphicx} 
\usepackage{epstopdf}
\usepackage[normalem]{ulem}
\usepackage{amsmath} 
\usepackage[pdftex]{hyperref}

\newcommand{\bra}[1]{\left\langle #1 \right|}
\newcommand{\ket}[1]{\left|#1\right\rangle}

\usepackage[dvipsnames]{xcolor}
\usepackage{soul}
\begin{document}

\title{Quantum Modeling of Scanning Near-Field Optical photons Scattered by an Atomic-Force Microscope Tip for Quantum Metrology}
\author{Soheil Khajavi$^{1}$, Zahra Shaterzadeh-Yazdi$^{2}$, Ali Eghrari$^{3}$, Mohammad Neshat$^{1}$\\
1.~School of Electrical and Computer Engineering, College of Engineering, University of Tehran, Tehran, Iran \\ 
2.~School of Engineering Science, College of Engineering, University of Tehran, Tehran, Iran\\
3.~Faculty of Natural, Mathematical $\&$ Engineering Sciences, King's College London, London, United Kingdom}
\date{\today}

\begin{abstract}
\noindent
Scattering scanning near-field optical microscopy (s-SNOM) is a promising technique for overcoming Abbe diffraction limit and substantially enhancing the spatial resolution in spectroscopic imaging. The s-SNOM works by exposing an atomic force microscope (AFM) tip to an optical electromagnetic~(EM) field, while the tip is so close to a dielectric sample that the incident beam lies within the near-field regime and displays nonlinear behaviour. We replace the incident EM field by photons generated by a single photon emitter, and propose a quantum model for the suggested system by employing electric-dipole approximation, image theory, and perturbation theory. Quantum state of scattered photons from the AFM tip is extracted from the proposed model, which contains information about electrical permittivity of the dielectric material beneath the tip. The permittivity of the sample can be extracted through spectroscopic setups. Our proposed scheme has potential applications for high-resolution quantum sensing and metrology, especially for quantum imaging and quantum spectroscopy.\\
\end{abstract}
\maketitle
\pagestyle{myheadings}
\markboth{Soheil Khajavi}{Quantum Sensing with Scanning Near-Field Optical photons Scattered by an Atomic-Force Microscope Tip}
\thispagestyle{empty}

\section{Introduction}

For the past few decades, various techniques have been proposed for overcoming the $\lambda/2$ diffraction limit, also known as the Abbe diffraction limit \cite{POHL1991243,durig1986near,harootunian1986super,betzig1987collection}. These techniques are based on apertures~\cite{betzig1991breaking} and scattering probes~\cite{betzig1987}. Scattering scanning near-field optical microscopy (s-SNOM) is one of the reliable techniques that has attracted more attention recently, due to the independence of resolution to the wavelength.
In this technique, the resolution only depends on the size of the setup.

The s-SNOM technique is leveraged on placing an atomic force microscope (AFM) tip close to the surface of a desired sample, with the tip exposed to a proper EM field. The incident EM field interacts with the AFM tip instead of directly interacting with the dielectric sample. In this approach, the resolution is limited to the size of the AFM tip. This technique provides imaging information with high nanoscale resolution, which has been recorded to be within the order of 1~nm~\cite{zenhausern1995}.

Several research groups have studied the interaction of a classical EM light with an AFM tip in spectroscopic setups, mainly through the image theory~\cite{porto2000,knoll2000,cvitkovic2007}. Also, interaction of a classical electromagnetic~(EM) light with an AFM tip doped with 3-level quantum particles has been studied recently~\cite{majorel2020}. To the best of our knowledge, interaction between photons incident to an AFM tip, in the presence of a dielectric material, has not yet been studied. 

\begin{figure}[ht] 
  \begin{center}
    \includegraphics[width=8cm]{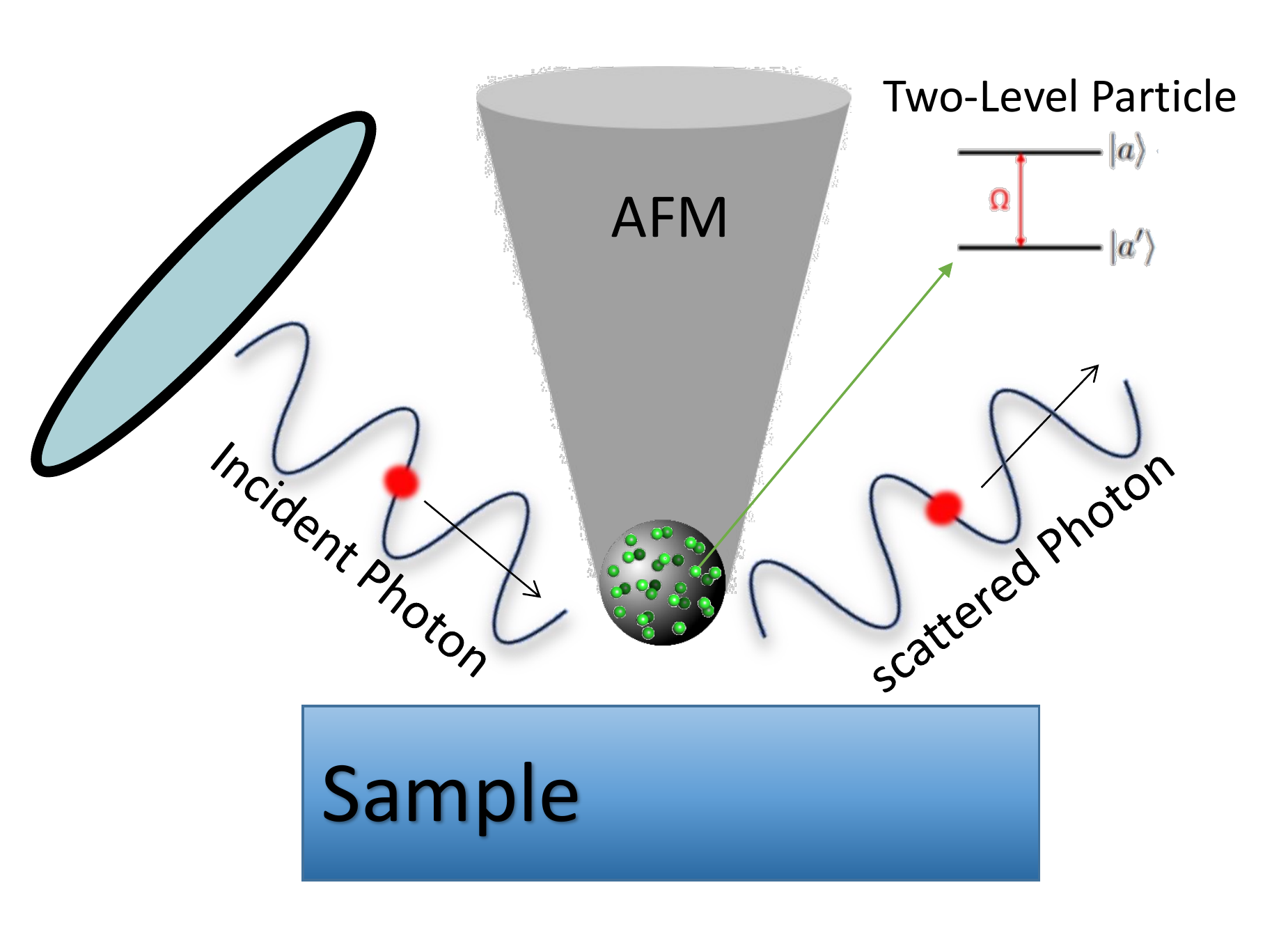}
    \caption{A simplified s-SNOM setup consisting of an incident quantized EM field, an AFM-tip doped with quantum dots, a nearby sample, and a scattered photon.}
    \label{fig1}
  \end{center}
\end{figure}
We investigate the interaction of near-field photons incident to an AFM tip, close to the surface of a dielectric sample. The goal is to study the effect of the dielectric material on the scattered photons from the AFM tip. As shown in Fig.~\ref{fig1}, we consider an AFM tip, possessing atomic-size quantum dots (QDs), such as phosphorous~(P) atoms doped in silicon~(Si)~\cite{dope1,dope2}, Si-based dangling-bonds~(DB)~\cite{sh1,sh2,sh3}, or diamond-based nitrogen-vacancy~(NV$^-$) centers~\cite{d1}. Incident photons interact with the tip and excite the quantum dots; as the QDs return to their ground state, photons are generated and scattered out of the tip. Since the AFM tip is placed close to the dielectric half-plane surface, spontaneous emission of photons from the AFM tip is modified at the presence of the dielectric surface, resulting in changes in the Fock state of the re-emitted photons.

To model the interaction between the AFM tip and the sample, we consider using the image theory. The AFM tip exposed by incident photons is modeled by an electric dipole. The sample is replaced by an effective image dipole containing the dielectric properties of the material, including the permittivity. The AFM tip is so close to the surface ($\approx30$~nm, the rim zone) that the near-field electrodynamics applies in this scale, and the van der Waals force is induced between the tip's dipole and its image dipole in the sample. The effect of the sample on the scattered photons from the AFM tip, is modeled by employing modified second-order perturbation theory. Fock state of the scattered photons, which is carrying information about the properties of the dielectric material, is then calculated. Electrical properties of the material can be extracted from a quantum-spectroscopy setup~\cite{mukai,lindner}.

%\textcolor{red}{The photons can be produced by single photon generators such as spontaneous parametric down-conversion (SPDC). These setups use the non-degenerate entangled photons with distinct frequencies in the visible and infrared spectrum. They detect the visible photons using visible detectors, while the interaction between the interacting photons and the sample is in infrared regime~\cite{mukai,lindner}}.

In the following sections, first we present the required information for modeling the proposed system. Then, quantum states of the desired system are obtained and the state of scattered photons from the AFM tip and the energy shift due to the presence of the sample are calculated. At the end, the concluding remarks are presented.

\section{Method}

In this section, we elaborate on our proposed system. The system is composed of an AFM tip, exposed to incident photons, that is in close proximity to a dielectric sample. The AFM tip is close enough to the sample that the van der Waals interaction dominates. In the following, detailed description for the components of the proposed model is presented. 

\subsection{AFM Tip and the Induced Image Dipole}

\begin{figure}[ht] 
  \begin{center}
    \includegraphics[width=8cm]{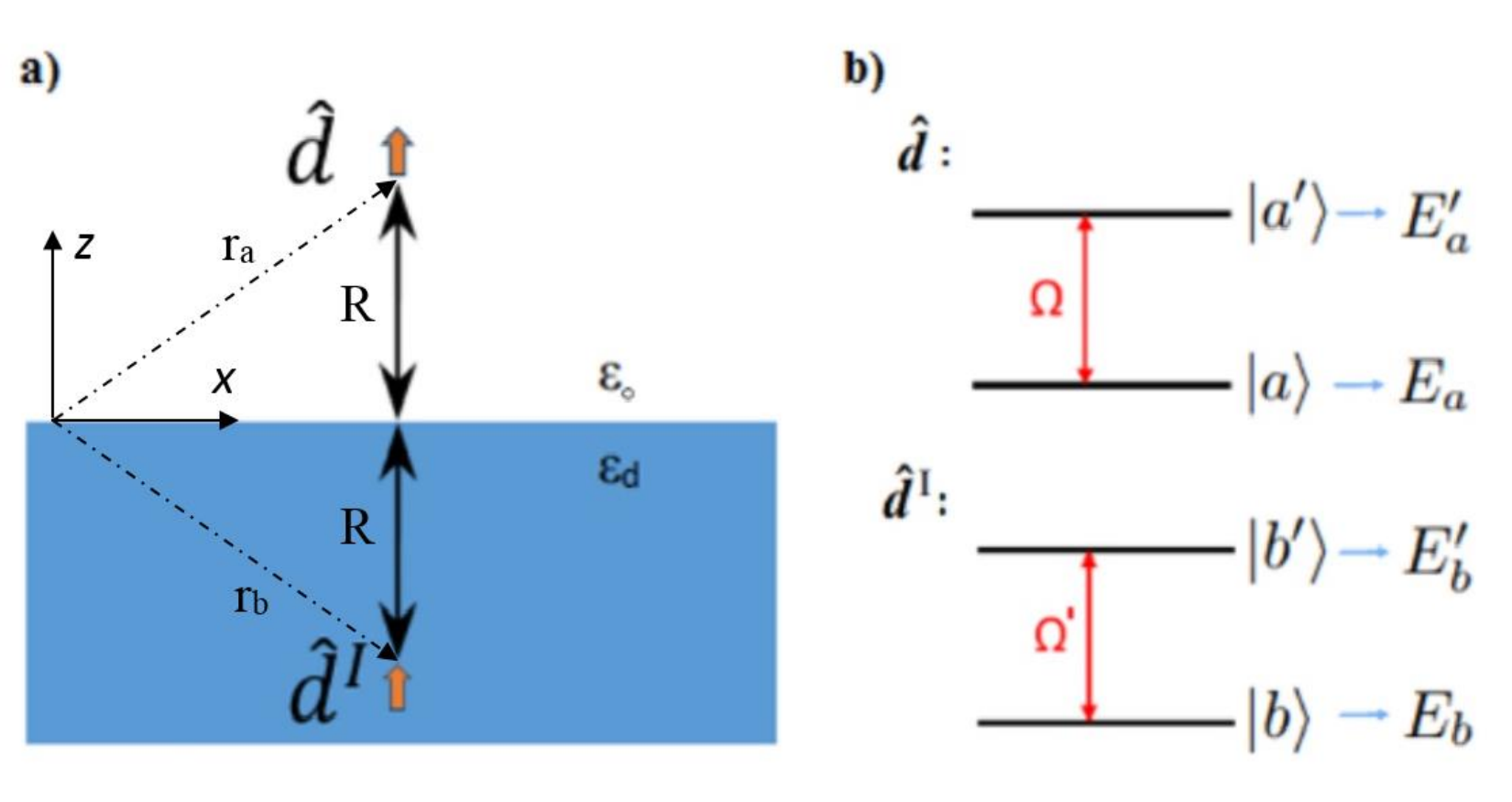}
    \caption{
a) Electric dipole moment, $\hat{d}$, representing the AFM tip exposed to incident photons, and its image dipole $\hat{d}^{I}$ in the dielectric material of the sample. b) The two-level dipole is represented by the two states $\ket{a}$ and $\ket{a'}$, with the corresponding energies $E_a$ and $E_a'$, respectively. The energy difference between the two states is represented by $\Omega$. The same notation applies for the image dipole, denoted by ``b".} 
     \label{fig2}
  \end{center}
 \end{figure}

As illustrated in Fig.~\ref{fig2}, an AFM tip exposed to incident photons can be modeled by a quantum electric dipole with the associated states given by $\ket{a}$ and $\ket{a'}$, and corresponding energies $E_a$ and $E'_a$, respectively. The energy difference between the two states, i.e.~$\Omega=E'_a-E_a$ is proportional to the frequency of the incident photons.
 
\begin{figure}[h!] 
  \begin{center}
    \includegraphics[width=4.5cm]{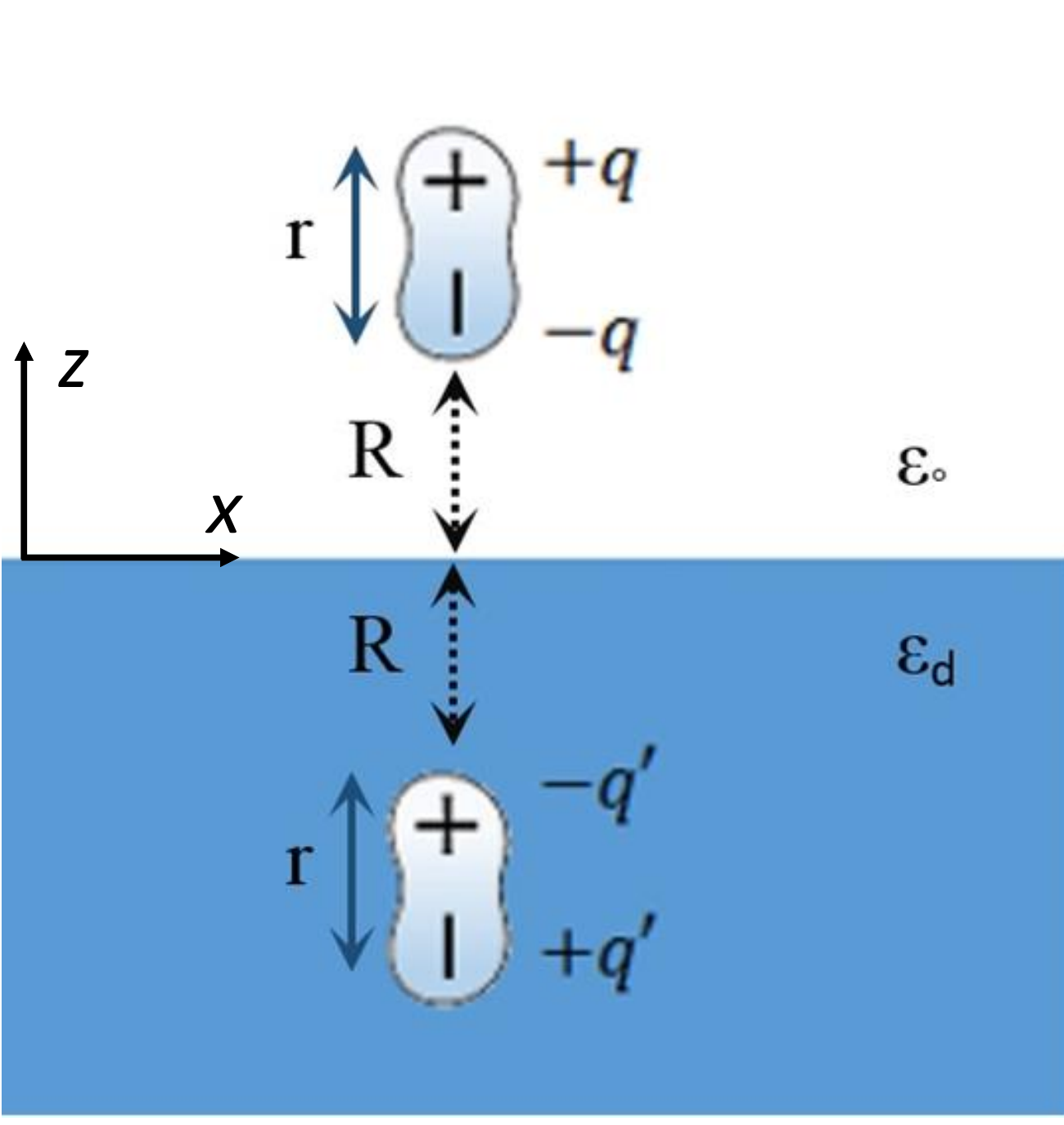}
    \caption{An electric dipole close to a dielectric material with dielectric constant $\epsilon_d$. The dipole charge distribution induces a dipole image with different charges within the dielectric material.}
    \label{fig3}
  \end{center}
\end{figure}

 Assuming that the tip's dipole is at a distance $R$ from the surface of the nearby dielectric sample, it results in an image dipole within the sample that is located at the same distance $R$ below the surface. According to the image theory, the whole effect of the half-space sample on the tip's dipole can be modeled by considering its corresponding image dipole. As shown in Fig.~\ref{fig3}, it is assumed that the charges of the image dipole are separated by the same distance $r$ as the main dipole, but with different charge values~\cite{jackson1999,hammond1960}. The image of a single charge in a dielectric material is given by
\begin{eqnarray} 
 q' &=& -\alpha q,\quad   \alpha =    \frac{\epsilon_d-1}{\epsilon_d+1}
   \label{2.2}
\end{eqnarray}
where $\alpha$ is the dielectric-air coefficient, which is a function of $\epsilon_d$, i.e.~the electric permittivity of the dielectric material. 

Many research groups have studied the coulomb energy of particle-surface interactions~\cite{mavroyannis1963,zaremba1976}. Subsequently, the energies associated with the states of the image dipole (i.e.~$\ket{b}$ and $\ket{b'}$) are given by,
\begin{eqnarray} 
E_b &=& \alpha^2 E_a, \quad
E_b' = \alpha^2 E_a',
   \label{2.4}
\end{eqnarray}
where $E_a$ and $E_a'$ are the energies associated with the states of the AFM tip's dipole. 
The same modification applies for the energy difference of the image dipole, i.e.~$\Omega'= \alpha^2 \Omega$. 

Due to van der Waals interaction between the tip's dipole and the image dipole, we expect to see a shift in the energy states of the tip's dipole, caused by the image dipole. The effect of this interaction will be presented in Sec.~\ref{vdW}. In the following section, the role of incident photons on the AFM tip is discussed.

\subsection{Photons Incident to the AFM Tip}

To model photons incident to the AFM tip, we consider the near-field interaction. Study of electromagnetic waves~(EM) and their quantum counterparts, i.e.~photons, in an area very close to a material lies in the near-field electrodynamics, which is a great field of interest ~\cite{pohl2012,girard1996,courtois1996, feron1993}. 

As photons are reflected from the AFM tip, there are a few of them with imaginary k-vectors, called the evanescent photons, that decay as they pass through the very small area between the AFM tip and the nearby dielectric sample. As a consequence, spontaneous emission of the photons scattered from the AFM tip is modified due to the nearby dielectric sample.
In other words, the presence of a dielectric material near to the AFM tip affects the incident photon and the tip interaction, resulting in a signature of the electric permittivity of the dielectric surface, in the quantum state of the photons scattered from the AFM tip.

Due to the near-field electrodynamics, interaction between the electric dipole of the tip and its image dipole is studied by the van der Waals force. This approach has many interest in scanning near-field optical microscopy~\cite{knoll2000}. To guarantee for being in the near-field region, hence ensuring the van der Waals interaction between the two dipoles, the distance between the AFM dipole and the corresponding image dipole should be small enough to satisfy,
\begin{eqnarray} 
 |r_a - r_b| &<<& c_0 \omega_{aa'}^{-1} ,  c_0\omega_{bb'}^{-1},  
    \label{3.1}
\end{eqnarray}
where $\omega_{aa'}^{-1} $ and $\omega_{bb'}^{-1}$ are the Bohr frequencies of the dipoles~\cite{keller2012}, $r_a$ and $r_b$ are the position vectors of the dipoles, and $c_0$ is the speed of light in vacuum, see Fig.~\ref{fig2}.

\subsection{Van der Waals Interaction Between the AFM Tip and the sample}\label{vdW}
To model the tip-sample van der Waals interaction, we propose a modified perturbed Hamiltonian representing the AFM tip's dipole and the image dipole interaction~\cite{feinberg1970}. The effective perturbation Hamiltonian corresponding to the instantaneous dipole–dipole interaction is obtained by \cite{keller2012},
\begin{eqnarray}
 \delta \hat{H} &=& -\frac{1}{|r_a-r_b|^3}\frac{1}{\epsilon_0} \hat{d}.\delta_T(\hat{r_a}-\hat{r_b}). \hat{d^I},
    \label{3.2}
\end{eqnarray}
where $\hat{d}$ and $\hat{d}^I$ are the electric-dipole moment operators of the main dipole and the image dipole, respectively~\cite{keller2012}. The parameter $\delta_T$ is the transverse delta function, representing the direction along the line passing through the dipoles.
If the dipoles are polarized in the $z$ direction, as shown in Fig.~\ref{fig2}, the inner product between the operators vanishes. The electric dipole operators are,
\begin{eqnarray} 
  \hat{d} &=& d_{aa'} (\hat{\sigma}_+^a + \hat{\sigma}_-^a)\;\; \text{and}\;\;
   \hat{d}^I = \alpha d_{bb'} (\hat{\sigma}_+^b + \hat{\sigma}_-^b)
    \label{3.3}
\end{eqnarray}
where $\alpha$ is given by Eq.~\ref{2.2} and contains information about the permittivity of the sample, $d_{ii'}$ for $i=\{a,b\}$ is the electric dipole moment, and $\hat{\sigma}_{\pm}^i$ are the ladder operators defined by $\hat{\sigma}_{+}^i=|i'\rangle\langle i|$ and $\hat{\sigma}_{-}^i=|i\rangle\langle i'|$.
Eventually, the perturbed Hamiltonian turns to
\begin{eqnarray} 
 \delta\hat{H} = & -\frac{1}{|r_a-r_b|^3}\frac{1}{\epsilon_0} \alpha  (\hat{\sigma}_+^a\hat{\sigma}_+^b +
 \nonumber\\&
 \hat{\sigma}_+^a\hat{\sigma}_-^b + \hat{\sigma}_-^a\hat{\sigma}_+^b +  \hat{\sigma}_-^a\hat{\sigma}_-^b).
     \label{3.4}
\end{eqnarray}
In the next section the effect of this perturbation on the quantum states of the photons, scattered from the AFM tip, is discussed.

\section{Quantum states and Energies of the System after Perturbation}

Quantum states of the proposed system are comprised of the quantum states of the tip's dipole, its image dipole in the dielectric sample, and the Fock state, $|n\rangle$, representing the photons incident on the AFM tip. These quantum states are given, in general, by
\begin{eqnarray} 
 \Sigma a_{1,n}\ket{a,b,n}, \Sigma a_{2,n}\ket{a,b',n}, \nonumber \\ \Sigma a_{3,n}\ket{a',b,n}, \Sigma a_{4,n}\ket{a',b',n}.
   \label{2.7}
\end{eqnarray}
In the following, we study the effect of perturbation $\delta\hat{H}$ on these states.
\subsection{Perturbed States}
Since perturbation in the system results from the effect of the image dipole on the tip's dipole, as presented in Eq.~\ref{3.2}, there is no \emph{direct} effect on the Fock state of the incident photons in the perturbation Hamiltonian. Furthermore, the first-order perturbation applies for charged particles, which is not the case for the dielectric sample, in our desired system. Therefore, modified second-order perturbation theory is implemented, and the resultant calculated perturbed states are given by,
\begin{eqnarray}
\sum_{n=1}^\infty
{\beta_{1n}\ket{a,b,n}^{(2)}} &=& \sum_{n=1}^\infty - \frac{1}{2(2R)^3}\frac{(a_{1n})^2(\frac{1}{\epsilon_0}\alpha)^2}{(\Omega(1+ \alpha^2))^2} \ket{a,b,n},
\nonumber\\
 
     \label{3.5}
\end{eqnarray}
where $\beta_{1n}$ is the coefficient of the perturbed state $\ket{a,b,n}$. Similar results can be obtained for other three states.

Presuming to have a condition under which single photons interact with the AFM tip (for instance, having a single-photon emitter as a source of generating the incident photons), the state of such photons would be represented by $\ket{1}$. Then, the density matrix of our desired system would be a $4\times4$ matrix given by, 
\begin{eqnarray} 
\rho_{ab} &=& \ket{\Psi}\bra{\Psi} =
 \begin{bmatrix}
    {\beta'_{1}}^2 & \beta'_{1}\beta'_{2} & \beta'_{1}\beta'_{3} & \beta'_{1}\beta'_{4} \\
    \beta'_{2}\beta'_{1} & {\beta'_{2}}^2 & \beta'_{2}\beta'_{3} & \beta'_{2}\beta'_{4} \\
     \beta'_{3}\beta'_{1} & \beta'_{3}\beta'_{2} & {\beta'_{3}}^3 & \beta'_{3}\beta'_{4}\\
      \beta'_{4}\beta'_{1} & \beta'_{4}\beta'_{2} & \beta'_{4}\beta'_{3} & {\beta'_{4}}^2  
 \end{bmatrix},
    \label{3.9}
\end{eqnarray}
where parameters $\beta'_j$ for $j\in\{1,2,3,4\}$ are the coefficients of every possible state in our system and are given by,
\begin{eqnarray} 
\beta'_1 = \beta'_3 = a_{1,3}- \frac{1}{2(2R)^3}\frac{{a_{1,3}}^2(\frac{1}{\epsilon_0}\alpha)^2}{(\Omega(1+\alpha^2))^2}, \nonumber\\
\beta'_2 = \beta'_4 = a_{2,4}- \frac{1}{2(2R)^3}\frac{{a_{2,4}}^2(\frac{1}{\epsilon_0}\alpha)^2}{(\Omega(1-\alpha^2))^2}.
    \label{3.10}
\end{eqnarray}

Suppose that using a proper mechanism, single photons scattered from the AFM tip, are detected by a single photon detector. These photons contain information about the electric properties of the dielectric sample. State of the scattered photons is obtained by reducing the state of the whole system, by tracing out the states of the subsystems associated with the AFM dipole and its image dipole interaction. Consequently, the state associated with the scattered photon is 
\begin{eqnarray} 
\ket{\Psi_1}&=&\sqrt{{\beta'_{1}}^2 +  {\beta'_{2}}^2 + {\beta'_{3}}^2 + {\beta'_{4}}^2} \ket{1}.
    \label{3.12}
\end{eqnarray}

Assuming that both dipoles are initially in their ground states,
then only $\beta'_1$ remains nonzero. Substituting its value from Eq.~\ref{3.10}, the state of the scattered photon is obtained to be,
\begin{eqnarray} 
\ket{\Psi_1} &=& (a_1- \frac{1}{2(2R)^3}\frac{a_1^2(\frac{1}{\epsilon_0}\alpha)^2}{(\Omega(1+\alpha^2))^2}) \ket{1},
    \label{3.13}
\end{eqnarray}
where $a_1$ is the coefficient of the unperturbed state $\ket{a,b,1}$. Equation~\ref{3.13} clearly shows that the state of a single scattered photon is bearing information about the properties of the dielectric sample, which is concealed in $\alpha$. Furthermore, the above equation demonstrates that the scattered-photon state depends on the distance $R$ of the tip from the sample, and the energy difference between the two states of the tip's dipole,~$\Omega$.

\subsection{Energy Shift due to Perturbation}

The effective perturbed Hamiltonian given by Eq.~\ref{3.2}, leads to an energy change, $\Delta E$, between the quantum states associated with the AFM tip, which is obtained by
\begin{eqnarray} 
\Delta E =-\frac{1}{(2R)^{3}} \frac{a_1^2(\frac{1}{\epsilon_0}\alpha)^2}{\Omega(1+\alpha^2)}.
    \label{3.14}
\end{eqnarray}
This energy change is negative, indicating that the van der Waals interaction reduces the energy difference between the two states of the tip's dipole, i.e.~$\Omega$. Also, the amount of $\Delta E$ varies with $R^{-3}$, which is denoting that the tip-sample force is inversely proportional to $R^{3}$. Consequently, the frequency of the photons scattered from the AFM tip is obtained by
\begin{eqnarray} 
\omega_s=\frac{\Omega-|\Delta E|}{\hbar}.
\end{eqnarray}
The photons' frequency contains information about the electrical properties of the dielectric material, which can be extracted from a quantum-spectroscopy setup~\cite{mukai,lindner}.

\section{Conclusion}

To conclude, we showed that the quantum behavior of an AFM tip exposed to incident photons is modified by existence of a dielectric material nearby. This modification is resulted by the van der Waals interaction between the tip's dipole and an its image dipole, representing the properties of the dielectric. Finally, the Fock state of a scattered single photon is obtained through the perturbation theory and can be used to specify the permittivity of the dielectric material.

The discussed scheme with classical EM field incident to the AFM tip, is the key component for scattering scanning near-field optical microscopy (s-SNOM), which is a promising technique for overcoming Abbe diffraction limit. We speculate that replacing the classical incident EM field in this scheme, by a stream of photons, substantially helps in enhancing the spatial resolution. Therefore, the proposed quantum s-SNOM provides the opportunity for quantum imaging, quantum spectroscopy, and quantum sensing with higher spatial resolution. 
\nocite{*}
\bibliography{refs}

\providecommand{\noopsort}[1]{}\providecommand{\singleletter}[1]{#1}%
\begin{thebibliography}{31}%
\makeatletter
\providecommand \@ifxundefined [1]{%
 \@ifx{#1\undefined}
}%
\providecommand \@ifnum [1]{%
 \ifnum #1\expandafter \@firstoftwo
 \else \expandafter \@secondoftwo
 \fi
}%
\providecommand \@ifx [1]{%
 \ifx #1\expandafter \@firstoftwo
 \else \expandafter \@secondoftwo
 \fi
}%
\providecommand \natexlab [1]{#1}%
\providecommand \enquote  [1]{``#1''}%
\providecommand \bibnamefont  [1]{#1}%
\providecommand \bibfnamefont [1]{#1}%
\providecommand \citenamefont [1]{#1}%
\providecommand \href@noop [0]{\@secondoftwo}%
\providecommand \href [0]{\begingroup \@sanitize@url \@href}%
\providecommand \@href[1]{\@@startlink{#1}\@@href}%
\providecommand \@@href[1]{\endgroup#1\@@endlink}%
\providecommand \@sanitize@url [0]{\catcode `\\12\catcode `\$12\catcode
  `\&12\catcode `\#12\catcode `\^12\catcode `\_12\catcode `\%12\relax}%
\providecommand \@@startlink[1]{}%
\providecommand \@@endlink[0]{}%
\providecommand \url  [0]{\begingroup\@sanitize@url \@url }%
\providecommand \@url [1]{\endgroup\@href {#1}{\urlprefix }}%
\providecommand \urlprefix  [0]{URL }%
\providecommand \Eprint [0]{\href }%
\providecommand \doibase [0]{https://doi.org/}%
\providecommand \selectlanguage [0]{\@gobble}%
\providecommand \bibinfo  [0]{\@secondoftwo}%
\providecommand \bibfield  [0]{\@secondoftwo}%
\providecommand \translation [1]{[#1]}%
\providecommand \BibitemOpen [0]{}%
\providecommand \bibitemStop [0]{}%
\providecommand \bibitemNoStop [0]{.\EOS\space}%
\providecommand \EOS [0]{\spacefactor3000\relax}%
\providecommand \BibitemShut  [1]{\csname bibitem#1\endcsname}%
\let\auto@bib@innerbib\@empty
%</preamble>
\bibitem [{\citenamefont {Pohl}(1991)}]{POHL1991243}%
  \BibitemOpen
  \bibfield  {author} {\bibinfo {author} {\bibfnamefont {D.}~\bibnamefont
  {Pohl}},\ }\bibfield  {title} {\bibinfo {title} {Scanning near-field optical
  microscopy (snom)}\ }(\bibinfo  {publisher} {Elsevier},\ \bibinfo {year}
  {1991})\ pp.\ \bibinfo {pages} {243--312}\BibitemShut {NoStop}%
\bibitem [{\citenamefont {D{\"u}rig}\ \emph {et~al.}(1986)\citenamefont
  {D{\"u}rig}, \citenamefont {Pohl},\ and\ \citenamefont
  {Rohner}}]{durig1986near}%
  \BibitemOpen
  \bibfield  {author} {\bibinfo {author} {\bibfnamefont {U.}~\bibnamefont
  {D{\"u}rig}}, \bibinfo {author} {\bibfnamefont {D.~W.}\ \bibnamefont
  {Pohl}},\ and\ \bibinfo {author} {\bibfnamefont {F.}~\bibnamefont {Rohner}},\
  }\bibfield  {title} {\bibinfo {title} {Near-field optical-scanning
  microscopy},\ }\href@noop {} {\bibfield  {journal} {\bibinfo  {journal}
  {Journal of applied physics}\ }\textbf {\bibinfo {volume} {59}},\ \bibinfo
  {pages} {3318} (\bibinfo {year} {1986})}\BibitemShut {NoStop}%
\bibitem [{\citenamefont {Harootunian}\ \emph {et~al.}(1986)\citenamefont
  {Harootunian}, \citenamefont {Betzig}, \citenamefont {Isaacson},\ and\
  \citenamefont {Lewis}}]{harootunian1986super}%
  \BibitemOpen
  \bibfield  {author} {\bibinfo {author} {\bibfnamefont {A.}~\bibnamefont
  {Harootunian}}, \bibinfo {author} {\bibfnamefont {E.}~\bibnamefont {Betzig}},
  \bibinfo {author} {\bibfnamefont {M.}~\bibnamefont {Isaacson}},\ and\
  \bibinfo {author} {\bibfnamefont {A.}~\bibnamefont {Lewis}},\ }\bibfield
  {title} {\bibinfo {title} {Super-resolution fluorescence near-field scanning
  optical microscopy},\ }\href@noop {} {\bibfield  {journal} {\bibinfo
  {journal} {Applied Physics Letters}\ }\textbf {\bibinfo {volume} {49}},\
  \bibinfo {pages} {674} (\bibinfo {year} {1986})}\BibitemShut {NoStop}%
\bibitem [{\citenamefont {Betzig}\ \emph
  {et~al.}(1987{\natexlab{a}})\citenamefont {Betzig}, \citenamefont
  {Isaacson},\ and\ \citenamefont {Lewis}}]{betzig1987collection}%
  \BibitemOpen
  \bibfield  {author} {\bibinfo {author} {\bibfnamefont {E.}~\bibnamefont
  {Betzig}}, \bibinfo {author} {\bibfnamefont {M.}~\bibnamefont {Isaacson}},\
  and\ \bibinfo {author} {\bibfnamefont {A.}~\bibnamefont {Lewis}},\ }\bibfield
   {title} {\bibinfo {title} {Collection mode near-field scanning optical
  microscopy},\ }\href@noop {} {\bibfield  {journal} {\bibinfo  {journal}
  {Applied physics letters}\ }\textbf {\bibinfo {volume} {51}},\ \bibinfo
  {pages} {2088} (\bibinfo {year} {1987}{\natexlab{a}})}\BibitemShut {NoStop}%
\bibitem [{\citenamefont {Betzig}\ \emph {et~al.}(1991)\citenamefont {Betzig},
  \citenamefont {Trautman}, \citenamefont {Harris}, \citenamefont {Weiner},\
  and\ \citenamefont {Kostelak}}]{betzig1991breaking}%
  \BibitemOpen
  \bibfield  {author} {\bibinfo {author} {\bibfnamefont {E.}~\bibnamefont
  {Betzig}}, \bibinfo {author} {\bibfnamefont {J.~K.}\ \bibnamefont
  {Trautman}}, \bibinfo {author} {\bibfnamefont {T.}~\bibnamefont {Harris}},
  \bibinfo {author} {\bibfnamefont {J.}~\bibnamefont {Weiner}},\ and\ \bibinfo
  {author} {\bibfnamefont {R.}~\bibnamefont {Kostelak}},\ }\bibfield  {title}
  {\bibinfo {title} {Breaking the diffraction barrier: optical microscopy on a
  nanometric scale},\ }\href@noop {} {\bibfield  {journal} {\bibinfo  {journal}
  {Science}\ }\textbf {\bibinfo {volume} {251}},\ \bibinfo {pages} {1468}
  (\bibinfo {year} {1991})}\BibitemShut {NoStop}%
\bibitem [{\citenamefont {Betzig}\ \emph
  {et~al.}(1987{\natexlab{b}})\citenamefont {Betzig}, \citenamefont
  {Isaacson},\ and\ \citenamefont {Lewis}}]{betzig1987}%
  \BibitemOpen
  \bibfield  {author} {\bibinfo {author} {\bibfnamefont {E.}~\bibnamefont
  {Betzig}}, \bibinfo {author} {\bibfnamefont {M.}~\bibnamefont {Isaacson}},\
  and\ \bibinfo {author} {\bibfnamefont {A.}~\bibnamefont {Lewis}},\ }\bibfield
   {title} {\bibinfo {title} {Collection mode near-field scanning optical
  microscopy},\ }\href@noop {} {\bibfield  {journal} {\bibinfo  {journal}
  {Applied physics letters}\ }\textbf {\bibinfo {volume} {51}},\ \bibinfo
  {pages} {2088} (\bibinfo {year} {1987}{\natexlab{b}})}\BibitemShut {NoStop}%
\bibitem [{\citenamefont {Zenhausern}\ \emph {et~al.}(1995)\citenamefont
  {Zenhausern}, \citenamefont {Martin},\ and\ \citenamefont
  {Wickramasinghe}}]{zenhausern1995}%
  \BibitemOpen
  \bibfield  {author} {\bibinfo {author} {\bibfnamefont {F.}~\bibnamefont
  {Zenhausern}}, \bibinfo {author} {\bibfnamefont {Y.}~\bibnamefont {Martin}},\
  and\ \bibinfo {author} {\bibfnamefont {H.}~\bibnamefont {Wickramasinghe}},\
  }\bibfield  {title} {\bibinfo {title} {Scanning interferometric apertureless
  microscopy: optical imaging at 10 angstrom resolution},\ }\href@noop {}
  {\bibfield  {journal} {\bibinfo  {journal} {Science}\ }\textbf {\bibinfo
  {volume} {269}},\ \bibinfo {pages} {1083} (\bibinfo {year}
  {1995})}\BibitemShut {NoStop}%
\bibitem [{\citenamefont {Porto}\ \emph {et~al.}(2000)\citenamefont {Porto},
  \citenamefont {Carminati},\ and\ \citenamefont {Greffet}}]{porto2000}%
  \BibitemOpen
  \bibfield  {author} {\bibinfo {author} {\bibfnamefont {J.}~\bibnamefont
  {Porto}}, \bibinfo {author} {\bibfnamefont {R.}~\bibnamefont {Carminati}},\
  and\ \bibinfo {author} {\bibfnamefont {J.-J.}\ \bibnamefont {Greffet}},\
  }\bibfield  {title} {\bibinfo {title} {Theory of electromagnetic field
  imaging and spectroscopy in scanning near-field optical microscopy},\
  }\href@noop {} {\bibfield  {journal} {\bibinfo  {journal} {Journal of Applied
  Physics}\ }\textbf {\bibinfo {volume} {88}},\ \bibinfo {pages} {4845}
  (\bibinfo {year} {2000})}\BibitemShut {NoStop}%
\bibitem [{\citenamefont {Knoll}\ and\ \citenamefont
  {Keilmann}(2000)}]{knoll2000}%
  \BibitemOpen
  \bibfield  {author} {\bibinfo {author} {\bibfnamefont {B.}~\bibnamefont
  {Knoll}}\ and\ \bibinfo {author} {\bibfnamefont {F.}~\bibnamefont
  {Keilmann}},\ }\bibfield  {title} {\bibinfo {title} {Enhanced dielectric
  contrast in scattering-type scanning near-field optical microscopy},\
  }\href@noop {} {\bibfield  {journal} {\bibinfo  {journal} {Optics
  communications}\ }\textbf {\bibinfo {volume} {182}},\ \bibinfo {pages} {321}
  (\bibinfo {year} {2000})}\BibitemShut {NoStop}%
\bibitem [{\citenamefont {Cvitkovic}\ \emph {et~al.}(2007)\citenamefont
  {Cvitkovic}, \citenamefont {Ocelic},\ and\ \citenamefont
  {Hillenbrand}}]{cvitkovic2007}%
  \BibitemOpen
  \bibfield  {author} {\bibinfo {author} {\bibfnamefont {A.}~\bibnamefont
  {Cvitkovic}}, \bibinfo {author} {\bibfnamefont {N.}~\bibnamefont {Ocelic}},\
  and\ \bibinfo {author} {\bibfnamefont {R.}~\bibnamefont {Hillenbrand}},\
  }\bibfield  {title} {\bibinfo {title} {Analytical model for quantitative
  prediction of material contrasts in scattering-type near-field optical
  microscopy},\ }\href@noop {} {\bibfield  {journal} {\bibinfo  {journal}
  {Optics express}\ }\textbf {\bibinfo {volume} {15}},\ \bibinfo {pages} {8550}
  (\bibinfo {year} {2007})}\BibitemShut {NoStop}%
\bibitem [{\citenamefont {Majorel}\ \emph {et~al.}(2020)\citenamefont
  {Majorel}, \citenamefont {Girard}, \citenamefont {Cuche}, \citenamefont
  {Arbouet},\ and\ \citenamefont {Wiecha}}]{majorel2020}%
  \BibitemOpen
  \bibfield  {author} {\bibinfo {author} {\bibfnamefont {C.}~\bibnamefont
  {Majorel}}, \bibinfo {author} {\bibfnamefont {C.}~\bibnamefont {Girard}},
  \bibinfo {author} {\bibfnamefont {A.}~\bibnamefont {Cuche}}, \bibinfo
  {author} {\bibfnamefont {A.}~\bibnamefont {Arbouet}},\ and\ \bibinfo {author}
  {\bibfnamefont {P.~R.}\ \bibnamefont {Wiecha}},\ }\bibfield  {title}
  {\bibinfo {title} {Quantum theory of near-field optical imaging with
  rare-earth atomic clusters},\ }\href@noop {} {\bibfield  {journal} {\bibinfo
  {journal} {JOSA B}\ }\textbf {\bibinfo {volume} {37}},\ \bibinfo {pages}
  {1474} (\bibinfo {year} {2020})}\BibitemShut {NoStop}%
\bibitem [{\citenamefont {Chan}\ \emph {et~al.}(2008)\citenamefont {Chan},
  \citenamefont {Tiago}, \citenamefont {Kaxiras},\ and\ \citenamefont
  {Chelikowsky}}]{dope1}%
  \BibitemOpen
  \bibfield  {author} {\bibinfo {author} {\bibfnamefont {T.-L.}\ \bibnamefont
  {Chan}}, \bibinfo {author} {\bibfnamefont {M.~L.}\ \bibnamefont {Tiago}},
  \bibinfo {author} {\bibfnamefont {E.}~\bibnamefont {Kaxiras}},\ and\ \bibinfo
  {author} {\bibfnamefont {J.~R.}\ \bibnamefont {Chelikowsky}},\ }\bibfield
  {title} {\bibinfo {title} {Size limits on doping phosphorus into silicon
  nanocrystals},\ }\href@noop {} {\bibfield  {journal} {\bibinfo  {journal}
  {Nano letters}\ }\textbf {\bibinfo {volume} {8}},\ \bibinfo {pages} {596}
  (\bibinfo {year} {2008})}\BibitemShut {NoStop}%
\bibitem [{\citenamefont {Melnikov}\ and\ \citenamefont
  {Chelikowsky}(2004)}]{dope2}%
  \BibitemOpen
  \bibfield  {author} {\bibinfo {author} {\bibfnamefont {D.~V.}\ \bibnamefont
  {Melnikov}}\ and\ \bibinfo {author} {\bibfnamefont {J.~R.}\ \bibnamefont
  {Chelikowsky}},\ }\bibfield  {title} {\bibinfo {title} {Quantum confinement
  in phosphorus-doped silicon nanocrystals},\ }\href@noop {} {\bibfield
  {journal} {\bibinfo  {journal} {Physical review letters}\ }\textbf {\bibinfo
  {volume} {92}},\ \bibinfo {pages} {046802} (\bibinfo {year}
  {2004})}\BibitemShut {NoStop}%
\bibitem [{\citenamefont {Livadaru}\ \emph {et~al.}(2010)\citenamefont
  {Livadaru}, \citenamefont {Xue}, \citenamefont {Shaterzadeh-Yazdi},
  \citenamefont {DiLabio}, \citenamefont {Mutus}, \citenamefont {Pitters},
  \citenamefont {Sanders},\ and\ \citenamefont {Wolkow}}]{sh1}%
  \BibitemOpen
  \bibfield  {author} {\bibinfo {author} {\bibfnamefont {L.}~\bibnamefont
  {Livadaru}}, \bibinfo {author} {\bibfnamefont {P.}~\bibnamefont {Xue}},
  \bibinfo {author} {\bibfnamefont {Z.}~\bibnamefont {Shaterzadeh-Yazdi}},
  \bibinfo {author} {\bibfnamefont {G.~A.}\ \bibnamefont {DiLabio}}, \bibinfo
  {author} {\bibfnamefont {J.}~\bibnamefont {Mutus}}, \bibinfo {author}
  {\bibfnamefont {J.~L.}\ \bibnamefont {Pitters}}, \bibinfo {author}
  {\bibfnamefont {B.~C.}\ \bibnamefont {Sanders}},\ and\ \bibinfo {author}
  {\bibfnamefont {R.~A.}\ \bibnamefont {Wolkow}},\ }\bibfield  {title}
  {\bibinfo {title} {Dangling-bond charge qubit on a silicon surface},\
  }\href@noop {} {\bibfield  {journal} {\bibinfo  {journal} {New Journal of
  Physics}\ }\textbf {\bibinfo {volume} {12}},\ \bibinfo {pages} {083018}
  (\bibinfo {year} {2010})}\BibitemShut {NoStop}%
\bibitem [{\citenamefont {Shaterzadeh-Yazdi}\ \emph {et~al.}(2014)\citenamefont
  {Shaterzadeh-Yazdi}, \citenamefont {Livadaru}, \citenamefont {Taucer},
  \citenamefont {Mutus}, \citenamefont {Pitters}, \citenamefont {Wolkow},\ and\
  \citenamefont {Sanders}}]{sh2}%
  \BibitemOpen
  \bibfield  {author} {\bibinfo {author} {\bibfnamefont {Z.}~\bibnamefont
  {Shaterzadeh-Yazdi}}, \bibinfo {author} {\bibfnamefont {L.}~\bibnamefont
  {Livadaru}}, \bibinfo {author} {\bibfnamefont {M.}~\bibnamefont {Taucer}},
  \bibinfo {author} {\bibfnamefont {J.}~\bibnamefont {Mutus}}, \bibinfo
  {author} {\bibfnamefont {J.}~\bibnamefont {Pitters}}, \bibinfo {author}
  {\bibfnamefont {R.~A.}\ \bibnamefont {Wolkow}},\ and\ \bibinfo {author}
  {\bibfnamefont {B.~C.}\ \bibnamefont {Sanders}},\ }\bibfield  {title}
  {\bibinfo {title} {Characterizing the rate and coherence of single-electron
  tunneling between two dangling bonds on the surface of silicon},\ }\href@noop
  {} {\bibfield  {journal} {\bibinfo  {journal} {Physical Review B}\ }\textbf
  {\bibinfo {volume} {89}},\ \bibinfo {pages} {035315} (\bibinfo {year}
  {2014})}\BibitemShut {NoStop}%
\bibitem [{\citenamefont {Shaterzadeh-Yazdi}\ \emph {et~al.}(2018)\citenamefont
  {Shaterzadeh-Yazdi}, \citenamefont {Sanders},\ and\ \citenamefont
  {DiLabio}}]{sh3}%
  \BibitemOpen
  \bibfield  {author} {\bibinfo {author} {\bibfnamefont {Z.}~\bibnamefont
  {Shaterzadeh-Yazdi}}, \bibinfo {author} {\bibfnamefont {B.~C.}\ \bibnamefont
  {Sanders}},\ and\ \bibinfo {author} {\bibfnamefont {G.~A.}\ \bibnamefont
  {DiLabio}},\ }\bibfield  {title} {\bibinfo {title} {Ab initio
  characterization of coupling strength for all types of dangling-bond pairs on
  the hydrogen-terminated si (100)-2$\times$ 1 surface},\ }\href@noop {}
  {\bibfield  {journal} {\bibinfo  {journal} {The Journal of Chemical Physics}\
  }\textbf {\bibinfo {volume} {148}},\ \bibinfo {pages} {154701} (\bibinfo
  {year} {2018})}\BibitemShut {NoStop}%
\bibitem [{\citenamefont {Schirhagl}\ \emph {et~al.}(2014)\citenamefont
  {Schirhagl}, \citenamefont {Chang}, \citenamefont {Loretz},\ and\
  \citenamefont {Degen}}]{d1}%
  \BibitemOpen
  \bibfield  {author} {\bibinfo {author} {\bibfnamefont {R.}~\bibnamefont
  {Schirhagl}}, \bibinfo {author} {\bibfnamefont {K.}~\bibnamefont {Chang}},
  \bibinfo {author} {\bibfnamefont {M.}~\bibnamefont {Loretz}},\ and\ \bibinfo
  {author} {\bibfnamefont {C.~L.}\ \bibnamefont {Degen}},\ }\bibfield  {title}
  {\bibinfo {title} {Nitrogen-vacancy centers in diamond: nanoscale sensors for
  physics and biology},\ }\href@noop {} {\bibfield  {journal} {\bibinfo
  {journal} {Annu. Rev. Phys. Chem}\ }\textbf {\bibinfo {volume} {65}},\
  \bibinfo {pages} {83} (\bibinfo {year} {2014})}\BibitemShut {NoStop}%
\bibitem [{\citenamefont {Mukai}\ \emph {et~al.}(2021)\citenamefont {Mukai},
  \citenamefont {Arahata}, \citenamefont {Tashima}, \citenamefont {Okamoto},\
  and\ \citenamefont {Takeuchi}}]{mukai}%
  \BibitemOpen
  \bibfield  {author} {\bibinfo {author} {\bibfnamefont {Y.}~\bibnamefont
  {Mukai}}, \bibinfo {author} {\bibfnamefont {M.}~\bibnamefont {Arahata}},
  \bibinfo {author} {\bibfnamefont {T.}~\bibnamefont {Tashima}}, \bibinfo
  {author} {\bibfnamefont {R.}~\bibnamefont {Okamoto}},\ and\ \bibinfo {author}
  {\bibfnamefont {S.}~\bibnamefont {Takeuchi}},\ }\bibfield  {title} {\bibinfo
  {title} {Quantum fourier-transform infrared spectroscopy for complex
  transmittance measurements},\ }\href@noop {} {\bibfield  {journal} {\bibinfo
  {journal} {Physical Review Applied}\ }\textbf {\bibinfo {volume} {15}},\
  \bibinfo {pages} {034019} (\bibinfo {year} {2021})}\BibitemShut {NoStop}%
\bibitem [{\citenamefont {Lindner}\ \emph {et~al.}(2020)\citenamefont
  {Lindner}, \citenamefont {Wolf}, \citenamefont {Kiessling},\ and\
  \citenamefont {K{\"u}hnemann}}]{lindner}%
  \BibitemOpen
  \bibfield  {author} {\bibinfo {author} {\bibfnamefont {C.}~\bibnamefont
  {Lindner}}, \bibinfo {author} {\bibfnamefont {S.}~\bibnamefont {Wolf}},
  \bibinfo {author} {\bibfnamefont {J.}~\bibnamefont {Kiessling}},\ and\
  \bibinfo {author} {\bibfnamefont {F.}~\bibnamefont {K{\"u}hnemann}},\
  }\bibfield  {title} {\bibinfo {title} {Fourier transform infrared
  spectroscopy with visible light},\ }\href@noop {} {\bibfield  {journal}
  {\bibinfo  {journal} {Optics express}\ }\textbf {\bibinfo {volume} {28}},\
  \bibinfo {pages} {4426} (\bibinfo {year} {2020})}\BibitemShut {NoStop}%
\bibitem [{\citenamefont {Jackson}(1999)}]{jackson1999}%
  \BibitemOpen
  \bibfield  {author} {\bibinfo {author} {\bibfnamefont {J.~D.}\ \bibnamefont
  {Jackson}},\ }\href@noop {} {\bibinfo {title} {Classical electrodynamics}}
  (\bibinfo {year} {1999})\BibitemShut {NoStop}%
\bibitem [{\citenamefont {Hammond}(1960)}]{hammond1960}%
  \BibitemOpen
  \bibfield  {author} {\bibinfo {author} {\bibfnamefont {P.}~\bibnamefont
  {Hammond}},\ }\bibfield  {title} {\bibinfo {title} {Electric and magnetic
  images},\ }\href@noop {} {\bibfield  {journal} {\bibinfo  {journal} {Proc.
  IEE}\ }\textbf {\bibinfo {volume} {107}},\ \bibinfo {pages} {306} (\bibinfo
  {year} {1960})}\BibitemShut {NoStop}%
\bibitem [{\citenamefont {Mavroyannis}(1963)}]{mavroyannis1963}%
  \BibitemOpen
  \bibfield  {author} {\bibinfo {author} {\bibfnamefont {C.}~\bibnamefont
  {Mavroyannis}},\ }\bibfield  {title} {\bibinfo {title} {The interaction of
  neutral molecules with dielectric surfaces},\ }\href@noop {} {\bibfield
  {journal} {\bibinfo  {journal} {Molecular Physics}\ }\textbf {\bibinfo
  {volume} {6}},\ \bibinfo {pages} {593} (\bibinfo {year} {1963})}\BibitemShut
  {NoStop}%
\bibitem [{\citenamefont {Zaremba}\ and\ \citenamefont
  {Kohn}(1976)}]{zaremba1976}%
  \BibitemOpen
  \bibfield  {author} {\bibinfo {author} {\bibfnamefont {E.}~\bibnamefont
  {Zaremba}}\ and\ \bibinfo {author} {\bibfnamefont {W.}~\bibnamefont {Kohn}},\
  }\bibfield  {title} {\bibinfo {title} {Van der waals interaction between an
  atom and a solid surface},\ }\href@noop {} {\bibfield  {journal} {\bibinfo
  {journal} {Physical Review B}\ }\textbf {\bibinfo {volume} {13}},\ \bibinfo
  {pages} {2270} (\bibinfo {year} {1976})}\BibitemShut {NoStop}%
\bibitem [{\citenamefont {Pohl}\ and\ \citenamefont
  {Courjon}(2012)}]{pohl2012}%
  \BibitemOpen
  \bibfield  {author} {\bibinfo {author} {\bibfnamefont {D.~W.}\ \bibnamefont
  {Pohl}}\ and\ \bibinfo {author} {\bibfnamefont {D.}~\bibnamefont {Courjon}},\
  }\href@noop {} {\emph {\bibinfo {title} {Near field optics}}},\ Vol.\
  \bibinfo {volume} {242}\ (\bibinfo  {publisher} {Springer Science \& Business
  Media},\ \bibinfo {year} {2012})\BibitemShut {NoStop}%
\bibitem [{\citenamefont {Girard}\ and\ \citenamefont
  {Dereux}(1996)}]{girard1996}%
  \BibitemOpen
  \bibfield  {author} {\bibinfo {author} {\bibfnamefont {C.}~\bibnamefont
  {Girard}}\ and\ \bibinfo {author} {\bibfnamefont {A.}~\bibnamefont
  {Dereux}},\ }\bibfield  {title} {\bibinfo {title} {Near-field optics
  theories},\ }\href@noop {} {\bibfield  {journal} {\bibinfo  {journal}
  {Reports on Progress in Physics}\ }\textbf {\bibinfo {volume} {59}},\
  \bibinfo {pages} {657} (\bibinfo {year} {1996})}\BibitemShut {NoStop}%
\bibitem [{\citenamefont {Courtois}\ \emph {et~al.}(1996)\citenamefont
  {Courtois}, \citenamefont {Courty},\ and\ \citenamefont
  {Mertz}}]{courtois1996}%
  \BibitemOpen
  \bibfield  {author} {\bibinfo {author} {\bibfnamefont {J.-Y.}\ \bibnamefont
  {Courtois}}, \bibinfo {author} {\bibfnamefont {J.-M.}\ \bibnamefont
  {Courty}},\ and\ \bibinfo {author} {\bibfnamefont {J.}~\bibnamefont
  {Mertz}},\ }\bibfield  {title} {\bibinfo {title} {Internal dynamics of
  multilevel atoms near a vacuum-dielectric interface},\ }\href@noop {}
  {\bibfield  {journal} {\bibinfo  {journal} {Physical Review A}\ }\textbf
  {\bibinfo {volume} {53}},\ \bibinfo {pages} {1862} (\bibinfo {year}
  {1996})}\BibitemShut {NoStop}%
\bibitem [{\citenamefont {Feron}\ \emph {et~al.}(1993)\citenamefont {Feron},
  \citenamefont {Reinhardt}, \citenamefont {Le~Boiteux}, \citenamefont
  {Gorceix}, \citenamefont {Baudon}, \citenamefont {Ducloy}, \citenamefont
  {Robert}, \citenamefont {Miniatura}, \citenamefont {Chormaic}, \citenamefont
  {Haberland} \emph {et~al.}}]{feron1993}%
  \BibitemOpen
  \bibfield  {author} {\bibinfo {author} {\bibfnamefont {S.}~\bibnamefont
  {Feron}}, \bibinfo {author} {\bibfnamefont {J.}~\bibnamefont {Reinhardt}},
  \bibinfo {author} {\bibfnamefont {S.}~\bibnamefont {Le~Boiteux}}, \bibinfo
  {author} {\bibfnamefont {O.}~\bibnamefont {Gorceix}}, \bibinfo {author}
  {\bibfnamefont {J.}~\bibnamefont {Baudon}}, \bibinfo {author} {\bibfnamefont
  {M.}~\bibnamefont {Ducloy}}, \bibinfo {author} {\bibfnamefont
  {J.}~\bibnamefont {Robert}}, \bibinfo {author} {\bibfnamefont
  {C.}~\bibnamefont {Miniatura}}, \bibinfo {author} {\bibfnamefont {S.~N.}\
  \bibnamefont {Chormaic}}, \bibinfo {author} {\bibfnamefont {H.}~\bibnamefont
  {Haberland}}, \emph {et~al.},\ }\bibfield  {title} {\bibinfo {title}
  {Reflection of metastable neon atoms by a surface plasmon wave},\ }\href@noop
  {} {\bibfield  {journal} {\bibinfo  {journal} {Optics communications}\
  }\textbf {\bibinfo {volume} {102}},\ \bibinfo {pages} {83} (\bibinfo {year}
  {1993})}\BibitemShut {NoStop}%
\bibitem [{\citenamefont {Keller}(2012)}]{keller2012}%
  \BibitemOpen
  \bibfield  {author} {\bibinfo {author} {\bibfnamefont {O.}~\bibnamefont
  {Keller}},\ }\href@noop {} {\emph {\bibinfo {title} {Quantum theory of
  near-field electrodynamics}}}\ (\bibinfo  {publisher} {Springer Science \&
  Business Media},\ \bibinfo {year} {2012})\BibitemShut {NoStop}%
\bibitem [{\citenamefont {Feinberg}\ and\ \citenamefont
  {Sucher}(1970)}]{feinberg1970}%
  \BibitemOpen
  \bibfield  {author} {\bibinfo {author} {\bibfnamefont {G.}~\bibnamefont
  {Feinberg}}\ and\ \bibinfo {author} {\bibfnamefont {J.}~\bibnamefont
  {Sucher}},\ }\bibfield  {title} {\bibinfo {title} {General theory of the van
  der waals interaction: A model-independent approach},\ }\href@noop {}
  {\bibfield  {journal} {\bibinfo  {journal} {Physical Review A}\ }\textbf
  {\bibinfo {volume} {2}},\ \bibinfo {pages} {2395} (\bibinfo {year}
  {1970})}\BibitemShut {NoStop}%
\bibitem [{\citenamefont {Bell}(2012)}]{bell2012}%
  \BibitemOpen
  \bibfield  {author} {\bibinfo {author} {\bibfnamefont {R.}~\bibnamefont
  {Bell}},\ }\href@noop {} {\emph {\bibinfo {title} {Introductory Fourier
  transform spectroscopy}}}\ (\bibinfo  {publisher} {Elsevier},\ \bibinfo
  {year} {2012})\BibitemShut {NoStop}%
\bibitem [{\citenamefont {Malherbe}\ \emph {et~al.}(2012)\citenamefont
  {Malherbe}, \citenamefont {Scott},\ and\ \citenamefont
  {Selinger}}]{malherbe2012}%
  \BibitemOpen
  \bibfield  {author} {\bibinfo {author} {\bibfnamefont {O.}~\bibnamefont
  {Malherbe}}, \bibinfo {author} {\bibfnamefont {P.~J.}\ \bibnamefont
  {Scott}},\ and\ \bibinfo {author} {\bibfnamefont {P.}~\bibnamefont
  {Selinger}},\ }\bibfield  {title} {\bibinfo {title} {Partially traced
  categories},\ }\href@noop {} {\bibfield  {journal} {\bibinfo  {journal}
  {Journal of Pure and Applied Algebra}\ }\textbf {\bibinfo {volume} {216}},\
  \bibinfo {pages} {2563} (\bibinfo {year} {2012})}\BibitemShut {NoStop}%
\end{thebibliography}%

\end{document}